\documentstyle[aps,prb,epsfig]{revtex}
\begin{document}
\twocolumn \wideabs{
\title{Landau Damping in a $2D$ Electron Gas
with Imposed Quantum Grid}
\author{I. Kuzmenko}
\address{Department of Physics,
Ben-Gurion University of the Negev, Beer-Sheva}
\date{\today}
\maketitle
\begin{abstract}
Dielectric properties of semiconductor substrate with imposed two
dimensional ($2D$) periodic grid of quantum wires or nanotubes
(quantum crossbars, QCB) are studied.  It is shown that a capacitive
contact between QCB and semiconductor substrate does not destroy the
Luttinger liquid character of the long wave QCB excitations.  However,
the dielectric losses of a substrate surface are drastically modified
due to diffraction processes on the QCB superlattice.  QCB-substrate
interaction results in additional Landau damping regions of the
substrate plasmons.  Their existence, form and the density of losses
are strongly sensitive to the QCB lattice constant.
\end{abstract}
 \vspace{\baselineskip}
 \noindent {P}{A}{C}{S}: 78.67.-n; 77.22.Gm; 73.90.+f
 \vspace{\baselineskip}
}
\section{Introduction}\label{sect-Introd}
Recently, new experimental techniques have made it possible to
grow nanostructures which are topologically one-dimensional
($1D$), such as quantum wires and carbon nanotubes. One of the
most exciting developments in this field is fabrication of $2D$
networks by means of self-assembling, etching, lithography and
imprinting techniques.\cite{Diehl,Wei} Arrays of interacting
quantum wires may be formed in organic materials and in striped
phases of doped transition metal oxides. Especially remarkable is
a recent experimental proposal to fabricate 2D periodic grids from
single-wall carbon nanotubes ({S}{W}{C}{N}{T}) suspended above a
dielectric substrate.\cite{Rueckes,Dalton} The possibility of
exciting a SWCNT by external electric field together with its
mechanical flexibility makes such a grid formed by nanotubes a
good candidate for an element of random access memory for
molecular computing. They are especially promising objects for
studying novel electronic correlation properties, which, in
particular, are relevant for tracing Luttinger liquid (LL)
finger-prints in two dimensions. This challenging idea is
motivated by noticing some unusual properties of electrons in Cu-O
planes in High-T$_c$ materials.\cite{Anders}

A double $2D$ grid, i.e., two superimposed crossing arrays of
parallel conducting quantum wires or nanotubes, represents a
specific nano-object -- quantum crossbars (QCB).  Its spectral
properties can not be treated in terms of purely 1D or 2D electron
liquid theory. A constituent element of QCB (quantum wire or
nanotube) possesses the LL-like spectrum.\cite{Bockrath,Egger} A
single array of parallel quantum wires is still a LL-like system
qualified as a sliding phase \cite{Mukho1} provided only the
electrostatic interaction between adjacent wires is taken into
account. If tunneling is suppressed and the two arrays are coupled
only by electrostatic interaction in the crosses, the system
possesses the LL zero energy fixed point, and besides, a rich Bose-type
excitation spectrum (plasmon modes) arises at finite energies in the
2D Brillouin zone (BZ).\cite{KGKA1,GKKA}

QCB is a unique system which possesses the properties of $1D$
and $2D$ liquid depending on the type of experimental probe. Some
possibilities of observation of $1D\to 2D$ crossover in transport
measurements were discussed in Ref.\cite{Mukho1} Several crossover
effects such as appearance of non-zero transverse space
correlations and periodic energy transfer between arrays ("Rabi
oscillations") were discussed in Refs.\cite{GKKA,KKGA,KGKA2}

The spectral peculiarities of QCB are directly manifested in its
interaction with an external ac electro-magnetic field or with
conducting and/or dielectric substrates, whose electrodynamic
properties are well known. To estimate this response one should
note that two main parameters characterizing the plasmon spectrum
in QCB are the Fermi velocity $v$ of electrons in a wire and the
QCB period $a$ (we assume both periods to be equal). These
parameters define both typical QCB plasmon wave numbers
$k=|{\bf{k}}|\sim Q=2\pi/a$ and typical plasmon frequencies
$\omega\sim \omega_{Q}=vQ$. Choosing according to
Refs.\cite{Egger,Rueckes} $v\approx8\cdot10^{7}$~cm/sec and
$a\approx20$~nm, one finds that characteristic plasmon frequencies
lie in far infrared region $\omega\sim 10^{14}$~sec$^{-1}$, while
characteristic wave vectors are estimated as
$q\sim10^{6}$~cm$^{-1}$.

In this paper, QCB interaction with semiconductor substrate is
studied.  Any surface wave excited in the substrate is coupled with
QCB-plasmon modes due to the substrate-QCB interaction.  This
interaction might be strong enough because surface plasmon waves exist
in the same frequency and wave vector area as plasmons in QCB (see
subsection \ref{subsect-H-Int} for details).  Therefore exciting the
substrate plasmons one can probe the QCB characteristics.  Indeed,
substrate-QCB interaction substantially changes the conventional
picture of substrate dielectric losses.  Due to such interaction, new
regions of Landau damping appear.  The existence of these regions
themselves, as well as their structure and the density of losses are
sensitive to both QCB period $a$ and the direction of the wave vector
${\bf{k}}$ of the initial wave.  Thus, dielectric losses in
QCB-substrate system serve as a good tool for studying QCB spectral
properties.

The structure of the paper is as follows. In Section
\ref{sect-H}, we briefly describe double square QCB interacting
with the dielectric substrate and introduce the necessary definitions.
Dielectric properties of the system considered are studied in
Section \ref{sect-DF}, where Dyson-type equations for the
polarization operator are obtained and analyzed. The detailed
description of new regions of Landau damping is presented in
Section \ref{sect-Damping}. In Conclusion we summarize
the results obtained.

\section{Quantum Crossbars on Semiconductor Substrate}\label{sect-H}
\subsection{Quantum Crossbars}\label{subsect-H-QCB}

A square QCB is a $2D$ grid, formed by two periodically crossed
perpendicular arrays of $1D$ quantum wires or carbon nanotubes. A
single wire is characterized by its radius $r_{0}$, length $L$,
and LL interaction parameter $g$.  The minimal nanotube radius is
$r_{0}\approx 0.35$~nm,\cite{Louie} maximal nanotube length is
$L\approx 1$~mm, and the LL parameter is estimated as $g\approx
0.3$.\cite{Egger}

\begin{figure}[htb]
\centering
\includegraphics[width=60mm,height=20mm,angle=0,]{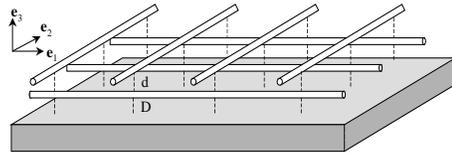}
\epsfxsize=70mm \caption{QCB on a substrate.  ${\bf{e}}_{\mu}$
($\mu=1,2,3$) are basic vectors of the coordinate system.  The vector
${\bf{e}}_{1}$ (${\bf{e}}_{2}$) is oriented along the first (second)
array.  The inter-array distance is $d$ and the distance between the
substrate and the first (lower) array is $D.$}
\label{Substrate}
\end{figure}

In experimentally realizable setups \cite{Rueckes} the QCB is a
cross-structure of suspended single-wall carbon nanotubes lying in two
parallel planes separated by an inter-plane distance $d,$ placed on a
dielectric or semiconductor substrate (see Fig.\ref{Substrate}).
Nevertheless, some generic properties of QCB may be described under
the assumption that QCB is a genuine $2D$ system.  We choose
coordinate system so that 1) the axes $x_{j}$ and corresponding basic
unit vectors ${\bf e}_{j}$ are oriented along the $j$-th array
($j=1,2$); 2)the $x_{3}$ axis is perpendicular to the QCB plane; 3)the
$x_{3}$ coordinate is zero for the second array, $-d$ for the first
one, and $-(d+D)$ for the substrate.  The basic vectors of the
reciprocal superlattice for a square QCB are $Q{\bf e}_{1,2},$
$Q=2\pi/a$ so that an arbitrary reciprocal superlattice vector
${\bf{m}}$ is a sum ${\bf{m}}={\bf{m}}_{1}+ {\bf{m}}_{2},$ where
${\bf{m}}_{j}=m_{j}Q {\bf{e}}_{j}$ ($m_j$ integer).  An arbitrary
vector ${\bf k}=k_{1}{\bf e}_{1}+k_{2}{\bf e}_{2}$ of reciprocal space
can be written as ${\bf q}+{\bf m}$ where ${\bf q}$ belongs to the
first BZ $|q_{1,2}|\leq Q/2.$

In a typical experimental setup\cite{Rueckes} the
characteristic lengths mentioned above have the following values
\begin{eqnarray*}
       d\approx 2 ~{\rm nm},\phantom{aa}L\approx 0.1 ~{\rm mm},
\end{eqnarray*}
so that the inequalities
\begin{eqnarray*}
    r_{0}\ll d\ll a\ll L
\end{eqnarray*}
are satisfied.

The QCB Hamiltonian,
\begin{equation}
    H_{QCB}=H_{1}+H_{2}+H_{12}
    \label{HamiltQCB}
\end{equation}
consists of three terms. The first (second) of them describes LL
in the first (second) array:
\begin{eqnarray*}
      H_j &=&
      \frac{\hbar}{2}\sum_{{\bf{qm}_j}}
      \bigg\{
           {v}{g}\pi_{j{\bf q}+{\bf m}_j}^{\dag}
           \pi_{j{\bf q}+{\bf m}_j}+\nonumber\\
      &&+
           \frac{\omega^2_{q_j+m_jQ}}{{v}{g}}
           \theta_{j,{\bf q}+{\bf m}_j}^{\dag}
           \theta_{j,{\bf q}+{\bf m}_j}
      \bigg\},\ \ \ j=1,2,
\end{eqnarray*}
where $(\theta_j,\pi_j)$ are the conventional canonically
conjugate boson fields and $\omega_k=v|k|$. The inter-array interaction,
\begin{eqnarray}
   &&H_{12}  =  \frac{\hbar\phi}{{v}{g}}\sum\limits_{{\bf{q,m}}}
     \xi_{q_1+m_1Q}\xi_{q_2+m_2Q}
     \theta_{1,{\bf q}+{\bf m}_1}^{\dag}
     \theta_{2,{\bf q}+{\bf m}_2},
     \label{Inter-k}
   \\
    &&\phi=\frac{gV_0r_0^2}{\hbar{v}{a}},
      \phantom{aa}
      \xi_{k}=\omega_{k}{\rm sign}k,
      \phantom{aa}
      V_{0}=\frac{2e^{2}}{d},
   \nonumber   
\end{eqnarray}
results from a short-range contact capacitive coupling in the
crosses of the bars.\cite{KGKA2} For {Q}{C}{B} formed by carbon
nanotubes, $\phi\sim0.007$.

The QCB Hamiltonian (\ref{HamiltQCB}) describes a system of
coupled harmonic oscillators, and can be diagonalized exactly. The
diagonalization procedure is, nevertheless, rather cumbersome due
to the mixing of states belonging to different bands and arrays.
However, separability of the interaction (\ref{Inter-k})
facilitates the diagonalization procedure and allows one to
describe the QCB spectrum analytically.\cite{KGKA2} Moreover,
being small, this interaction grossly conserves the unperturbed
$1D$ systematics of levels and states, at least in the low energy
region corresponding to the first few energy bands. This means
that perturbed eigenstates could be described in terms of the same
quantum numbers (array number, band number and quasimomentum) as
the unperturbed eigenstates of an ``empty'' lattice ($\phi=0$).

Such a description fails in two specific regions of the reciprocal
space ${\bf k}$.  The first one is the vicinity of the high symmetry
lines $k_{j}=nQ/2$ with $n$ integer (the lines with $n=\pm 1$ include
BZ boundaries).  Around these lines, the {\em inter-band} mixing is
significant and two modes from adjacent bands are degenerate.  The
second region is the vicinity of the resonant lines $k_{1}\pm
k_{2}=nQ$ where the eigenfrequencies of the unperturbed plasmons
$$\omega_{j}({\bf{k}})=v|k_j|, \ \ \ j=1,2,$$
from the same band propagating along two arrays coincide,
$\omega_{1}({\bf{k}})=\omega_{2}({\bf{k+m}})$. Around resonant
lines, {\em inter-array} mixing is significant and at these lines
two modes corresponding to different arrays are degenerate. In
what follows we restrict our consideration by the first BZ
quarter, $0\le{q_1,q_2}\le{Q}/{2}$. Here the dispersion laws
outside the inter-band and inter-array mixing regions conserve
their unperturbed form.

\subsection{Substrate characteristics}\label{subsect-H-Sub}

The substrate is described by the Hamiltonian
\begin{equation}
 H_s=H_K+H_C,
 \label{Hsub}
\end{equation}
where
\begin{eqnarray*}
  H_{K} =
  \sum_{\bf{qm}}\varepsilon_{\bf{q+m}}c_{\bf{q+m}}^{\dag}c_{\bf{q+m}},
  \ \ \ \
  \varepsilon_{{\bf{k}}}=\displaystyle{\frac{\hbar^{2}k^{2}}{2m}},
\end{eqnarray*}
is a kinetic energy of the substrate electrons with effective mass $m$
and quadratic dispersion law (we omit the irrelevant spin variables),
and
\begin{eqnarray*}
  H_C &=& \frac{1}{2}\sum_{\bf{qm}}U_{\bf{q+m}}
          \rho_{\bf{q+m}}^{\dag}\rho_{\bf{q+m}},
  \\
\rho_{\bf{k}}&=&\frac{1}{L}\sum_{\bf{k'}}c_{\bf{k'}}^{\dag}c_{\bf{k+k'}},
  \phantom{a}U_{\bf{k}}=\frac{2\pi e^2}{k},
  \phantom{a}{\bf{k}}={\bf{q}}+{\bf{m}},
\end{eqnarray*}
is Coulomb interaction within the substrate.

Dielectric properties of substrate {\it per se} are described by
its dielectric function $\epsilon_{s}({\bf{k}},\omega)$,
\begin{equation}
 \frac{1}{\epsilon_{s}({\bf{k}},\omega)}=
 1+U_{\bf{k}}\Pi_{s}({\bf k},\omega).
 \label{DielFunSub}
\end{equation}
Within the RPA approach, the polarization operator
$\Pi_{s}({\bf{k}},\omega)$ is presented by the Lindhard expression
\begin{eqnarray}
  \left(\Pi_{s}({\bf k},\omega)\right)^{-1} &=&
    \left(\Pi_{0}({\bf k},\omega)\right)^{-1}-U_{\bf k}
    \label{Pi_s},
\end{eqnarray}
with
\begin{eqnarray}
  \Pi_{0}({\bf k},\omega) &=&
    \frac{1}{L^2}\sum\limits_{\bf k'}
    \frac{\vartheta(\varepsilon_F-\varepsilon_{\bf k'})
         -\vartheta(\varepsilon_F-\varepsilon_{\bf k+k'})}
         {\hbar\omega-(\varepsilon_{\bf k+k'}-
                     \varepsilon_{\bf k'})+i0}.
 \label{Pi_0}
\end{eqnarray}

\begin{figure}[h]
\centerline{\epsfig{figure=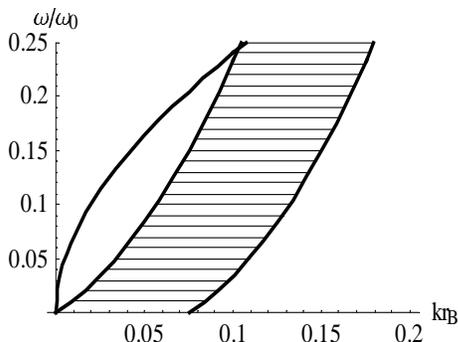,width=60mm,height=45mm,angle=0}}
\caption{Dispersion of the substrate plasmons (upper line) and
quasi-continuum spectrum of electron-hole excitations, (dashed
area). Frequency is measured in $\omega_0=v_F/r_B$ units.}
 \label{Disp0}
\end{figure}

Active branches of substrate excitations are the surface density
fluctuations which consist of $2D$ electron-hole pairs and surface
plasmons with dispersion law \cite{Stern}
\begin{eqnarray*}
 \omega_{s}({\bf{k}})=v_F k\sqrt{1+\frac{1}{2 kr_B}},
 \ \ \
 r_B=\displaystyle{\frac{\hbar^2}{me^2}},
 \ \ \
 k=|{\bf{k}}|.
\end{eqnarray*}
The {R}{P}{A} spectrum of surface excitations is shown in
Fig.\ref{Disp0}.

In the case of {Ga}{As}, the substrate parameters are
$m=0.068m_0$, $m_0$ is the free electron mass,
$v_F=8.2\cdot{10}^6$~cm/sec, $r_B=0.78$~nm,
$\omega_0=v_F/r_B={1.05}\times{10}^{14}$~sec$^{-1}$. For
$k<k^{*}\approx 0.104r_B^{-1}$, the plasmon frequency lies above
the continuum spectrum of electron-hole pairs and the substrate
plasmons are stable. Besides, one may easily satisfy the resonance
condition for the collective plasmon mode near the stability
threshold $k\sim k^{*}$ and QCB excitations with frequency
$\omega\sim\omega^{*}=2.6\cdot{10}^{13}$~sec$^{-1}$. For large
enough $k>k^{*}$ the plasmon dispersion curve lies within the
quasi-continuum spectrum and plasmons become unstable with respect
to decay into electron-hole pairs (Landau damping of the substrate
plasmons). Dielectric losses of an isolated substrate are
described by an imaginary part $\Im\epsilon_{s}({\bf{k}},\omega)$
of its dielectric function (\ref{DielFunSub}). This imaginary part
is non-zero within the dashed region in Fig.\ref{Disp0} due to
appearance of imaginary part
\begin{eqnarray*}
  \Im\Pi_0({\bf k},\omega) &&=
  \frac{m}{2\pi\hbar^2}
  \frac{1}{\kappa^2}
  \left[
      \vartheta(\kappa^2-\nu_{+}^2)
       \sqrt{\kappa^2-\nu_{+}^2}-
  \right.\nonumber\\&&\left.-
      \vartheta(\kappa^2-\nu_{-}^2)
       \sqrt{\kappa^2-\nu_{-}^2}
  \right],
  \\
  \kappa=\frac{k}{k_{F}}, && \ \ \ \nu_{\pm}=\nu\pm\kappa^{2}/2,
  \ \ \ \nu=\frac{\omega}{v_{F}k_{F}}
\end{eqnarray*}
of the bare polarization operator $\Pi_0({\bf k},\omega)$
(\ref{Pi_0}).

\subsection{Interaction}\label{subsect-H-Int}

Interaction between {Q}{C}{B} and substrate is a capacitive coupling
of charge fluctuations in the substrate with collective modes in the
quantum wires.  Assuming the distance $D$ between the first array and
the substrate to be much smaller than the distance $d$ between arrays,
one can keep only the interaction $H_{s1}$ between the substrate and
the first array.  The interaction between the substrate density
fluctuation at the point ${\bf r}\equiv (x_1,x_2)$ and the density
fluctuations located in the vicinity of the point $x'_1$ which belongs
to the $n_2$-th wire of the first array is described by its amplitude
$W(x_1-x'_1,x_2-n_2a)$, where
\begin{eqnarray*}
    W({\bf{r}}) &=&
    \frac{\sqrt{2}e^2
          \vartheta
          \left(
               1-{\displaystyle{\frac{|x_1|}{r_0}}}
          \right)}
         {\sqrt{\left|{\bf{r}}\right|^2+D^2}}.
\end{eqnarray*}
The theta function in the numerator describes the screening of
Coulomb interaction within a wire or nanotube.\cite{Sasaki} In the
momentum representation the interaction Hamiltonian between
substrate and array has the form:
\begin{equation}
  H_{s1}= \sqrt{\frac{\hbar}{vg}}\sum_{\bf{mq}}W_{{\bf{q}}+{\bf{m}}}
             \rho_{\bf{q+m}}\theta_{1,{\bf{q+m}}_1}^{\dag},
  \label{Hs1}
\end{equation}
where
\begin{eqnarray}
  W_{{\bf{k}}} = ik_1\sqrt{\frac{v g}{\hbar a}}\int d{\bf r}
  W({\bf{r}})e^{i{\bf{kr}}}
  \label{W_k}
\end{eqnarray}
is proportional to a Fourier component of the interaction amplitude
$W({\bf{r}})$.

Finally, the Hamiltonian of QCB interacting with a semiconductor
substrate is a sum of Hamiltonians (\ref{HamiltQCB}), (\ref{Hsub})
and (\ref{Hs1}),
\begin{equation}
  H = H_{QCB}+H_s+H_{s1}.
  \label{H}
\end{equation}

\section{Dielectric Function}\label{sect-DF}

High frequency properties of the system at zero temperature are
determined by zeroes of its dielectric function
\begin{equation}
 \frac{1}{\epsilon({\bf{k}},\omega)}=
 1+U_{\bf{k}}\Pi({\bf k},\omega).
 \label{DielFun}
\end{equation}
Here
\begin{eqnarray*}
  \Pi({\bf{k}},\omega)=-\frac{i}{\hbar}\int\limits_{0}^{\infty}dt
  e^{i\omega t}
         \left\langle\left[
              \rho_{\bf{k}}(t),
              \rho_{\bf{k}}^{\dag}(0)
         \right]\right\rangle,
\end{eqnarray*}
is the polarization of the {\it substrate interacting with QCB},
$\rho_{\bf{k}}(t)=e^{iHt/\hbar}\rho_{\bf{k}}e^{-iHt/\hbar}$ is the
density of the {\it substrate} electrons in the Heisenberg
representation, and averaging is performed over the ground state
of the Hamiltonian (\ref{H}).

The Umklapp processes stimulated by the interaction between the
substrate and the first array (\ref{Hs1}) as well as the interaction
between arrays (\ref{Inter-k}), produce modes with wave vectors
${\bf{q}}+{\bf{m}}$ with various inverse lattice vectors
${\bf{m}}$.  This necessarily leads to appearance of non-diagonal
polarization operators
\begin{eqnarray*}
  &&\Pi({\bf q} +{\bf m},{\bf q} +{\bf m}';\omega) =
  \nonumber\\&&=
  -\frac{i}{\hbar}\int\limits_{0}^{\infty}dt
  e^{i\omega t}
  \big\langle\big[
             \rho_{\bf{q+m}}(t)
             \rho_{\bf{q+m'}}^{\dag}(0)
  \big]\big\rangle.
\end{eqnarray*}
In what follows we always consider a fixed frequency $\omega$ and
a fixed wave vector ${\bf q}$ from the BZ. So the variables
${\bf{q}}$ and $\omega$ are omitted below for simplicity. In the
framework of {R}{P}{A} approach, $\Pi({\bf{m}},{\bf{m}}')$
satisfies the Dyson-type equation
\begin{eqnarray}
  \Pi({\bf{m,m}}') &=& \Pi_{s}({\bf{m}})\delta_{{\bf{m,m}}'}+
  \Pi_{s}({\bf{m}})W_{\bf{m}}\Xi_1(m_1,{\bf{m}}').
  \label{EqDyson1}
\end{eqnarray}
The first term $\Pi_s({\bf{m}})$ in the right hand side is the
substrate polarization (\ref{Pi_s}) of the isolated substrate
itself, ${W}_{\bf{m}}\equiv{W}_{{\bf{q}}+{\bf{m}}}$ is a bare
vertex (\ref{W_k}) which describes substrate - (first) array
interaction, and
\begin{eqnarray}
  \Xi_j(m_j,{\bf{m}}') &=&
   -\frac{i}{\hbar}\int\limits_{0}^{\infty}dt
  e^{i\omega t}
  \big\langle\big[
       \theta_{j,{\bf{q+m}}_j}(t),
       \rho_{{\bf{q}}+{\bf{m}}'}^{\dag}(0)
  \big]\big\rangle
  \label{Dj}
\end{eqnarray}
is the correlation function of the $j$-{t}{h} array mode and the
substrate plasmon.

The Dyson equation (\ref{EqDyson1}) should be completed by two
equations for the correlation functions (\ref{Dj}) ($j=1,2$)
\begin{eqnarray}
  \Xi_1(m_1,{\bf{m}}') &=&
   D_1^0(m_1)\sum_{m_2}W_{\bf{m}}\Pi({\bf{m,m}}')+
  \nonumber\\&+&
   D_1^0(m_1)\sum_{m_2}\Phi_{\bf{m}}
  \Xi_2(m_2,{\bf{m}}'),
  \label{EqDyson2}\\
  \Xi_2(m_2,{\bf{m}}') &=&
   D_2^0(m_2)\sum_{m_1}\Phi_{\bf{m}}
   \Xi_1(m_1,{\bf{m}}').
  \label{EqDyson3}
\end{eqnarray}
Here $D_{j}^{0}(m_j)$ ($j=1,2$) is the bare correlation function
of the $j$-{t}{h} array modes
\begin{eqnarray*}
  D_{j}^{0}(m_j) &=&
  -\frac{i}{v g}\int\limits_{0}^{\infty}dt
  e^{i\omega t}\left\langle\left[
                    \theta_{j,{\bf{q+m}}_j}(t),
                    \theta_{j,{\bf{q+m}}_j}^{\dag}(0)
               \right]\right\rangle_{0}
  \nonumber\\&=&
  \frac{1}{\omega^2-v^2(q_j+m_j)^2},
\end{eqnarray*}
and another bare vertex
$$
\Phi_{\bf{m}}=\frac{\hbar\phi}{{v}{g}}{\xi}_{q_1+m_1Q}{\xi}_{q_2+m_2Q},
$$
describes the separable inter-array interaction (\ref{Inter-k}).

Solving the system of equations (\ref{EqDyson1}), (\ref{EqDyson2})
and (\ref{EqDyson3}) one obtains the diagonal element
$\Pi({\bf m})\equiv \Pi({\bf m,m})$ of the polarization operator
\begin{equation}
  \left[\Pi({\bf m})\right]^{-1}=
  \left[\Pi_s({\bf m})\right]^{-1}-
        |W_{\bf{m}}|^2D({\bf m}).
  \label{Polar-Solution}
\end{equation}
The second term on the right-hand side of this equation describes
renormalization of the substrate polarization operator
$\Pi_s({\bf{m}})$ by interaction between the substrate and QCB.
The factor $D({\bf m})$ is a renormalized correlation function
of modes of the first array
\begin{eqnarray}
  \left[D({\bf m})\right]^{-1}&=&
  \left[D_1^0(m_1)\right]^{-1}
  -\left(w({\bf m})+\varphi(m_1)\right).
  \label{Dm}
\end{eqnarray}
The first term $w({\bf{m}})$ describes the effective interaction
between the first array and substrate
\begin{eqnarray}
  w({\bf{m}}) &=&F(m_{1})-
  |W_{\bf{m}}|^2\Pi_s({\bf m}),\nonumber\\
  F(m_{1})&=&\sum_{m_{2}}|W_{\bf{m}}|^2\Pi_s({\bf m}).
       \label{w}
\end{eqnarray}
The second one $\varphi(m_1)$ is the effective interaction between
arrays
\begin{eqnarray}
  \frac{\omega_{m_{1}}^{2}}{\varphi(m_1)} &=&
   \left[
    \phi^{2}\sum\limits_{m_2}\omega_{m_{2}}^{2}D_2^0(m_2)
   \right]^{-1}-\Psi_{m_{1}},
  \label{phi} \\
  \omega_{m_{j}}&=&v|q_{j}+m_{j}Q|,
  \nonumber
\end{eqnarray}
renormalized by Coulomb interaction of array modes with the
substrate plasmons,
\begin{eqnarray}
   \Psi_{m_1} &=&
         \sum_{{m'_1}\ne{m_1}}
         \frac{\omega_{m'_{1}}^{2}}
         {\left(D_1^0(m'_1)
         \right)^{-1}-F(m_1')}.
  \label{M}
\end{eqnarray}
Equations (\ref{Polar-Solution}) - (\ref{M}) together with
definition (\ref{DielFun}) solve the problem of dielectric
properties of the combined system QCB-substrate.

The spectrum of collective excitations in QCB-substrate system is
determined by zeros of the dielectric function
$\epsilon({\bf{q}},\omega)=0.$ The key question here is the
robustness of the QCB spectrum against interaction with $2D$
substrate excitations. Detailed analysis shows that in the long wave
limit $q\ll Q$ the interaction just renormalizes the bare
dispersion laws of the arrays, conserving its LL linearity. This
result verifies stability of QCB plasmons with respect to
substrate-QCB interaction.

The QCB-substrate interaction also results in occurrence of some
special lines in the BZ. These lines correspond to resonant
interaction of the substrate with the first or the second array.
The resonance condition $\omega_{s}({\bf{k}})=\omega_j({\bf{k}})$
is fulfilled along the line $LJIN$ for $j=1$ and along the line
$KBM$ for $j=2$ in Fig.\ref{30nm} below.

\section{Landau Damping}\label{sect-Damping}

As was mentioned in subsection\ref{subsect-H-Sub} above, dielectric
losses of an isolated substrate are related to Landau damping due to
decay of substrate plasmons with momentum $k>k^{*}\approx
0.104r_B^{-1}$ into electron-hole pair.  The substrate-QCB interaction
remarkably modifies the conventional picture of substrate plasmon
dielectric losses.  Due to QCB-substrate interaction, new domains of
Landau damping appear in addition to the dashed region in
Fig.\ref{Disp0}.  Indeed, outside the initial instability region where
$\Im\epsilon_{s}({\bf{k}},\omega)=0$, nonzero imaginary part
$\Im\epsilon({\bf{k}},\omega)$ (\ref{DielFun}) exists if the imaginary
part of the bare polarization operator
$\Im\Pi_0({\bf{k}}+{\bf{m}},\omega)$ differs from zero at least for
one of the reciprocal lattice vectors ${\bf{m}}$.  The main
contribution to $\Im\epsilon({\bf{k}},\omega)$ is related to the
renormalization term $w({\bf m})$ in Eq.(\ref{w}) due to Umklapp
processes along the $x_{2}$ axis (summation over $m_{2}$ in the
expression for the function $F(m_{1})$ is implied).  It is
proportional to the fourth power of QCB - substrate interaction
$W^{4}$.  The Umklapp processes along both directions $x_{1,2}$
contribute also to the renormalization term $\varphi(m_{1})$ in
Eq.(\ref{phi}).  However, they contain additional small parameter
$\phi^{4}$ related to inter-array interaction within QCB. These terms
are not taken into account.  Thus, the possible Umklapp vectors
have the form ${\bf m}_{2}=m_{2}Q{\bf e}_{2},$ $m_{2}=\pm 1,\pm 2,
\ldots$ and in what follows we will label them by an integer number
$m_{2}.$

\begin{figure}[h]
\centerline{\epsfig{figure=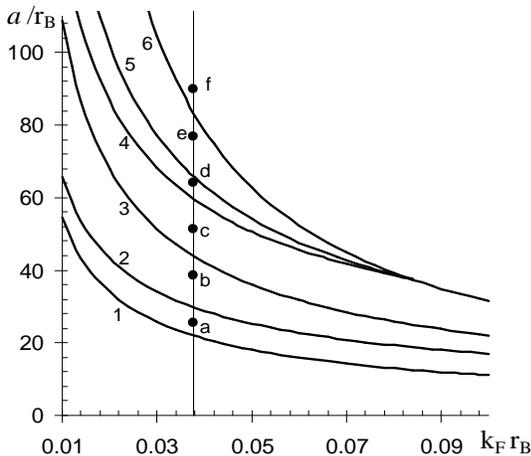,width=70mm,height=60mm,
angle=0}}
\caption{Phase diagram describing appearance and structure of
new regions of Landau damping. Lines $1-6$ separate different types
of new damping regions. Points $a-f$ corresponds to the structures
displayed below in Figs. \ref{20nm}-\ref{70nm}.}
\label{diagram}
\end{figure}

The structure of the new Landau damping regions and their existence
itself is governed by interplay between the Fermi momentum of the
substrate $k_{F}$ and the {Q}{C}{B} period $a$.  The first of these
parameters defines the width of the two particle excitation band
(dashed region in Fig.\ref{Disp0}) while the second determines the
minimal reciprocal vector $Q$.  In the case of sufficiently thick QCB
superlattice (small $a$) and sufficiently low electron density within
the substrate (small $k_{F}$), Umklapp processes are always
ineffective because they change an initial plasmon wavevector into the
outer part of the instability region.  This means that only plasmons
with momenta $|k|>k^{*}$ decay into electron-hole pairs.

Increasing the QCB period or the Fermi momentum turn the Umklapp
processes effective and additional Landau damping regions appear
within the circle $|k|\leq k^{*}$.  The first factors involved are
the smallest Umklapp vectors $\pm 1,$ then new damping regions
appear corresponding to the Umklapp vectors $\pm 2$ and so on.  As
a result one gets a rich variety of possible damping scenarios. We
describe them with the help of a ``phase diagram'' in the
$a-k_{F}$ plane displayed in Fig.\ref{diagram} (actually
dimensionless coordinates $a/r_{B}$ and $k_{F}r_{B}$ are used).
Here the set of curves labelled by numbers $1-6$ separate the
regions of parameters corresponding to the different Umklapp
vectors and different structures of the new damping regions. There
are no additional Landau damping regions below the first line.
Above the sixth line Landau damping takes place within the whole
circle $k\leq k^{*}.$ Above lines with numbers $2n-1,$ the Umklapp
vector $\pm n$ becomes effective.  Corresponding additional
damping region first has the form of a tail touching the initial
Landau damping region $|k|\geq k^{*}$.  This tail turns to the
additional damping band well separated from the initial one above
the line number $2n$ ($n<3$) within some sector of directions in
$k$-space (in what follows, these directions will be labelled by
corresponding arcs of the circle).

\begin{figure}[h]
\centerline{\epsfig{figure=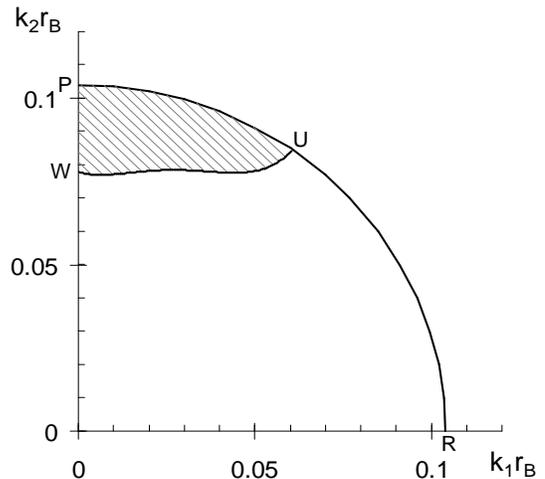,width=70mm,height=63mm,
angle=0}} \caption{New Landau damping region $PUW$ for QCB with
period $a=20$~nm (point $a$ in Fig.\ref{diagram}) corresponds to
the Umklapp vector $-1.$} \label{20nm}
\end{figure}

\begin{figure}[h]
\centerline{\epsfig{figure=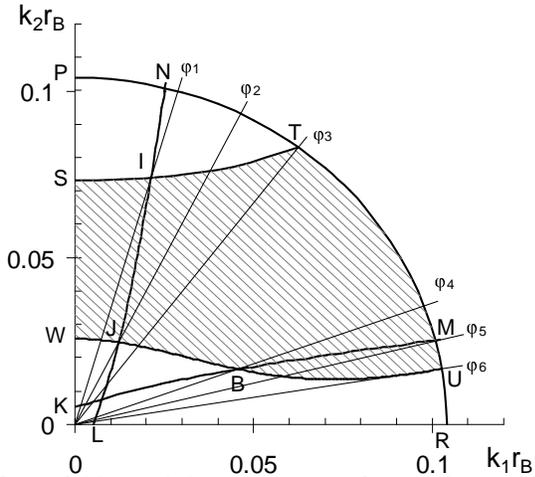,width=70mm,height=63mm,
angle=0}}
\caption{Additional damping region $STUW$ for $a=30$~nm
(point $b$ in Fig.\ref{diagram}) corresponds to the Umklapp vector
$-1$ and describes Landau damping tail (within the arc $TU$) or
separate landau band (within the arc $TP$).  Other details of this
figure are explained in the text.}
\label{30nm}
\end{figure}

\begin{figure}[h]
\centerline{\epsfig{figure=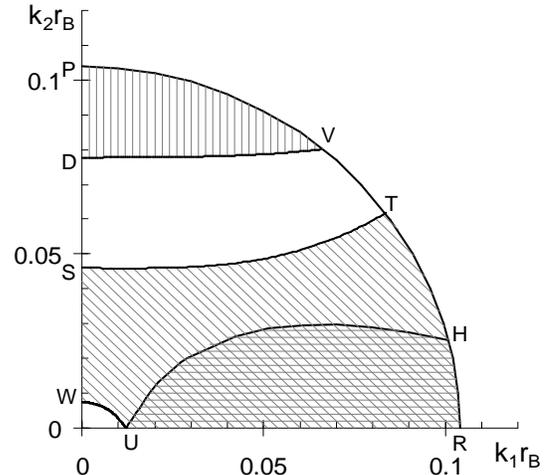,width=70mm,height=63mm,
angle=0}}
\caption{In the case $a=40$~nm (point $c$ in Fig.\ref{diagram}),
new Landau damping regions $PVD,$ $STRUW,$ and $UHR$
correspond to the Umklapp vectors $-2,$ $-1,$ and $+1,$.}
\label{40nm}
\end{figure}

\begin{figure}[h]
\centerline{\epsfig{figure=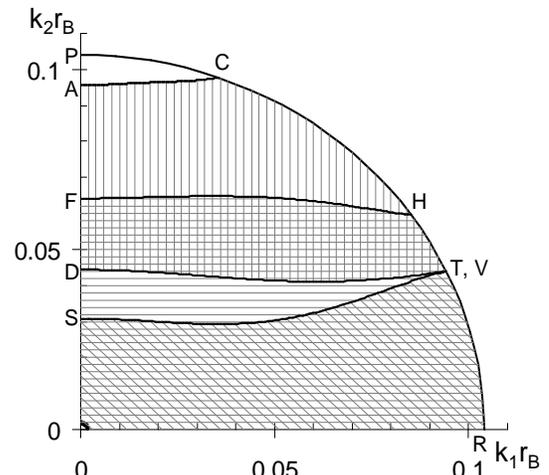,width=70mm,height=63mm,
angle=0}} \caption{New regions of Landau damping for $a=50$~nm
(point $d$ in Fig.\ref{diagram}).  Regions $ACVD$, $STR,$ and
$FHR$ correspond to the Umklapp vectors $-2,$ $-1,$ and $+1$}
\label{50nm}
\end{figure}

\begin{figure}[h]
\centerline{\epsfig{figure=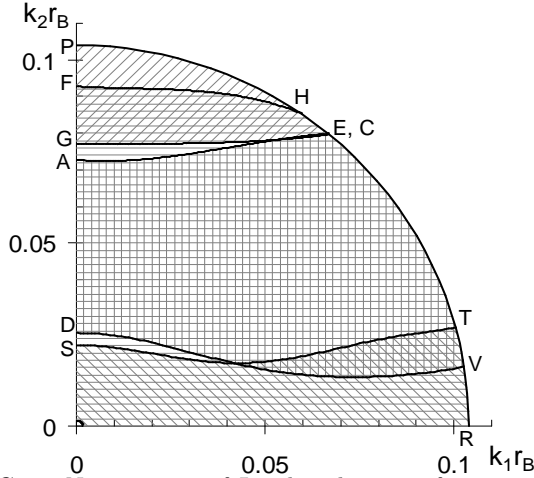,width=70mm,height=63mm,
angle=0}}
\caption{New regions of Landau damping for $a=60$~nm (point
$e$ in Fig.\ref{diagram}).  Besides the regions $ACVD,$ $STR,$ and
$FHR$ corresponding, as in the case $a=50nm,$ to the Umklapp vectors
$-2,$ $-1,$ and $+1,$ new Umklapp vector $-3$ appears (region
$PEG$).}
\label{60nm}
\end{figure}

\begin{figure}[h]
\centerline{\epsfig{figure=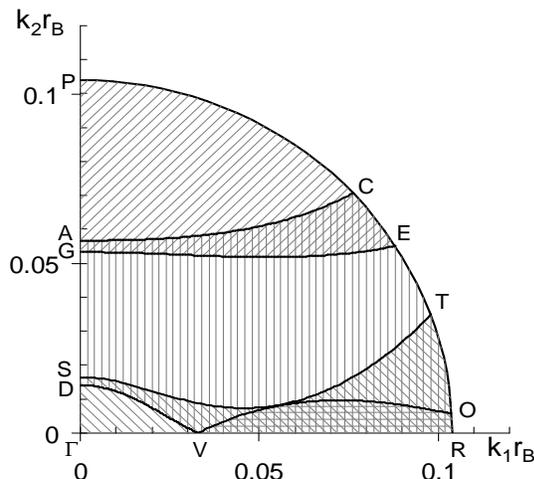,width=70mm,height=63mm,
angle=0}}
\caption{The case $a=70$~nm (point $f$ in Fig.\ref{diagram}).
New regions of Landau damping $PEG,$ $ACVD,$ $STR,$
whole sector $\Gamma PR,$ and $VOR$ correspond to the Umklapp vectors $-3,$
$-2,$ $-1,$ $+1,$ and $+2.$}
\label{70nm}
\end{figure}

Possible structures of new damping regions corresponding to some
representative points $a-f$ in the $a-k_{F}$ plane (see
Fig.\ref{diagram}) are displayed in detail in
Figs.\ref{20nm}-\ref{70nm}.  All these figures correspond to the
$GaAs$ value of $k_{F}r_{B}\approx 0.038.$ Generally speaking we
should display new damping region within the whole circle $|k|\leq
k^{*}$ in the plane $k_{1},k_{2}.$ The circle center $\Gamma$ is
placed at the origin (we did not put the letter $\Gamma$ in
Figs.\ref{20nm}-\ref{60nm}).  But this region is always symmetric with
respect to reflection $k_{1}\to -k_{1}$ and with respect to the
combined reflection $k_{2}\to -k_{2}, m_{2}\to -m_{2}.$ This enables
us to describe the damping scenarios within the quarter $k_{1,2}\geq
0$ only (complete picture of the new damping region can be easily
obtained from the displayed one with the help of the
reflections mentioned above).

Damping of the substrate plasmon occurs inside the arc $PR$ in
Figs.\ref{20nm}-\ref{70nm} when at least one of the points in the phase
space with coordinates $({\bf{k}}+{\bf{m}}_2,\omega_s(k))$ lies within
the quasi-continuum spectrum of the electron-hole excitations, whereas
the ``mother'' point $({\bf{k}},\omega_s(k))$ lies above the continuum
(above the dashed area in Fig.\ref{Disp0}).  As was mentioned above,
for small enough {Q}{C}{B} period $a<a_1=17.3$~nm, the basic
reciprocal lattice vector $Q{\bf{e}}_2$ is too large, the points
$({\bf{k}}+{\bf{m}}_2,\omega_s(k))$ lie outside the quasi-continuum
for all $m_2$ and additional Landau damping region does not exist.  It
appears only for $a>a_1$ ($Q<2k^{*}+2k_F$).  For $a_1<a<a_2=23.6$~nm
($2k^{*}+2k_F>Q>2k^{*}$), this is the region $PUW$ (see
Fig.\ref{20nm}) corresponding to the $m_{2}=-1$ (in all
Figs.\ref{20nm}-\ref{70nm}, the regions related to this Umklapp vector
are always hatched by the hatching tilted to left).  As a result, the
damping tails touching the initial Landau damping region appear in
certain directions of the ${\bf{k}}$ plane.

For $a_2<a<a_3=34.6$~nm ($2k^{*}>Q>k^{*}+k_F$), new damping region is
related to the same Umklapp vector $-1$, but now it has a
strip-like structure bounded by the line $STUW$ in Fig.\ref{30nm}.
Note that the damping is absent within the region $PTS$.  As a result,
it is possible to divide the angle region $0\le\varphi\le\pi/2$ into
three sectors.  Within the first one $PT$, $0\le\varphi\le\varphi_3$,
a new damping region is separated from the initial one.  The second
sector $TU$, $\varphi_3\le\varphi\le\varphi_6$, corresponds to a new
damping tail.  Finally, within the third sector $UR$ ,
$\varphi_6\le\varphi\le{\pi}/{2}$, new damping region does not exist
at all.

For larger QCB period, $a_3<a<a_4=47.1$~nm ($k^{*}+k_F>Q>k^{*}$), the
new damping regions have a more complicated structure.  In fact the
damping area consists of three parts (see Fig.\ref{40nm}).  The first
one, $STRUW$ corresponds to the Umklapp vector $-1.$ Note that it
is shifted to the bottom with respect to the previous case $a=30$~nm.
This part overlaps with the second part $HRU$.  The latter corresponds
to the Umklapp vector $+1$ and is hatched in Figs.\ref{40nm}-\ref{60nm}
by horizontal hatching.  The second region $PDV$
corresponds to the Umklapp vector $-2$ (in Figs.\ref{40nm}-\ref{70nm}
such a regions are always hatched by vertical
hatching).  As a result in the direction close enough to the $k_{2}$
axis, one gets a new damping tail with $m_{2}=-2$) and well separated
new damping band with Umklapp vector $-1.$

Further increase of QCB period $a_4<a<a_5=52.3$~nm,
$k^{*}>Q>2(k^{*}+k_F)/3$ ($k^{*}\approx 2.8 k_{F}$ for $GaAs$) leads
to further extension of new damping regions.  The region $ADVC$
corresponding to the Umklapp vector $-2$ is partially separated from
the initial damping region.  It overlaps with the region $FHR\Gamma$
($m_{2}=+1$) which in its turn overlaps with the region $STR\Gamma$
($m_{2}=-1$).  Actually the two latter regions do not include an
extremely small vicinity of the origin $\Gamma$ which is not shown in
Fig.\ref{50nm}.  Visible coincidence of the points $V$ and $T$ in
Fig.\ref{50nm} is an artefact of the accuracy of the figure.

Within the next interval of QCB periods $a_5<a<a_6=65.2$~nm
($2(k^{*}+k_F)/3>Q>2k_F)$) new damping region $GPE$ corresponding to
the Umklapp vector $-3$ appears.  This region is hatched in
Figs.\ref{60nm},\ref{70nm} by the hatching tilted to the right.  Beside
that the regions $ADVC$ ($m_{2}=-2$), $FHR\Gamma$ ($+1$), and
$STR\Gamma$ ($-1$) are present.  As in the previous figure, visible
coincidence of the points $E$ and $C$ in Fig.\ref{60nm} is an
artefact of the accuracy of the figure.

Finally for $a>a_6$ ($Q<2k_F$), Landau damping emerges in the whole
circle $|k|\leq k^{*}.$ This occurs due to processes with
Umklapp vector $+1.$ The corresponding damping region covers the whole
quarter.  Therefore we did not hatched it at all and used the same
horizontal hatching for the new region $VOR$ corresponding to the
Umklapp vector $+2.$ The region $DACRV$ is related to the
Umklapp vector $-2.$ We emphasize that the vertex $V$ of this region
at the same time is the vertex of the region $VOR,$ this is not an
accidental approximate coincidence as in the two previous figures.  The
regions $GPE$ and $\Gamma STR $ are related to the Umklapp vectors $-3$
and $-1$ correspondingly.

\begin{figure}[h]
\centerline{\epsfig{figure=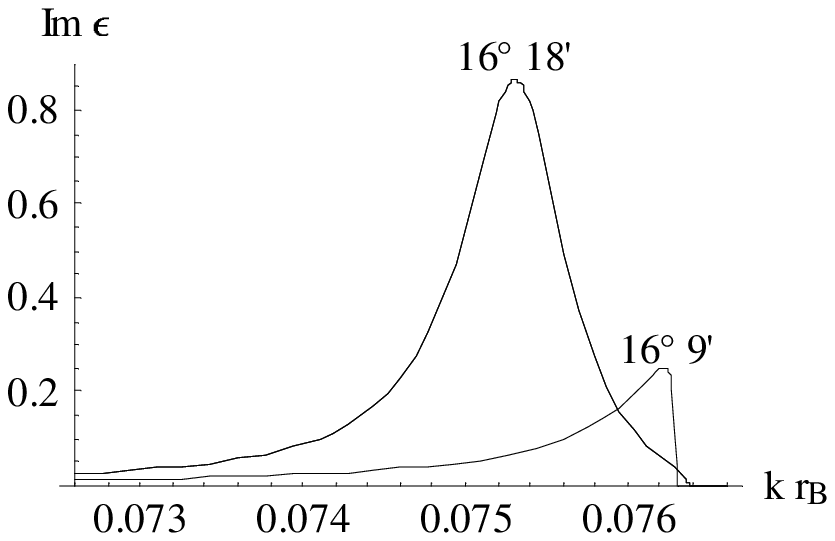,width=70mm,height=50mm,
angle=0}}
\caption{Damping tail for $\varphi\approx16^{\circ}$. Precursor of
the resonant peak and the resonant peak are resolved quite well.}
\label{ImDFAngl-16}
\end{figure}

Thus, the general structure of the additional damping regions is
described in Figs.\ref{20nm}-\ref{70nm}.  However there is an
additional structure of these regions.  This fine structure
is related to possible resonance interaction between the substrate
plasmons and the QCB plasmons of the first or second array.  The
resonance condition for the first (second) array is written as
$\omega_{s}({\bf{k}})=\omega_{1}({\bf{k}})$
($\omega_{s}({\bf{k}})=\omega_{2}({\bf{k}})$).

\begin{figure}[h]
\centerline{\epsfig{figure=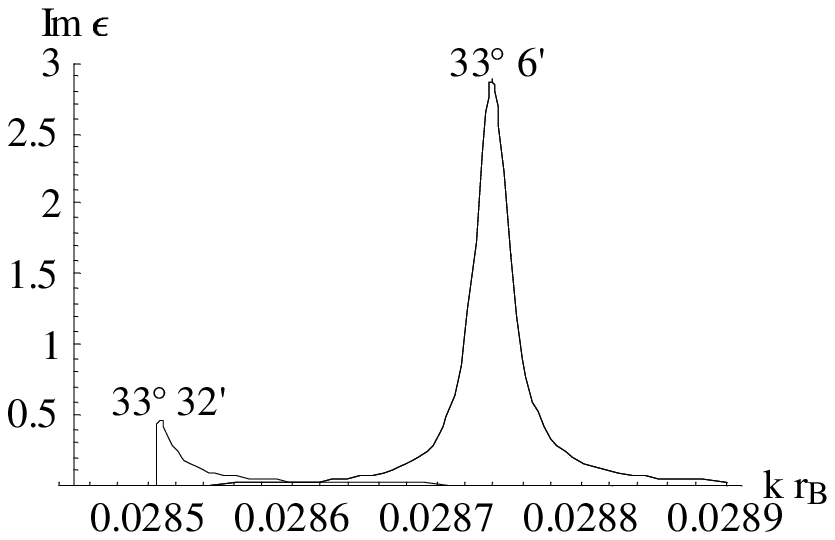,width=70mm,height=50mm,
angle=0}} \caption{Damping tail for $\varphi\approx33^{\circ}$.
Precursor of the resonant peak and the resonant peak are resolved
quite well.} \label{ImDFAngl-33}
\end{figure}

Consider such fine structure of new Landau damping region in
details for QCB with period $a=30$~nm. Here the resonance
conditions are satisfied along the lines $LJIN$ and $KBM$
(Fig.\ref{30nm}). These lines intersect with the damping region
boundaries at the points $J,I,B$ and $MB$ which define four rays
$OI$, $OJ$, $OB$, and $OU$ and four corresponding angles
$\varphi_{1}\approx16^{\circ}11',$
$\varphi_{2}\approx33^{\circ}19',$ $\varphi_{4}\approx69^{\circ},$
$\varphi_{5}\approx76^{\circ}.$ The resonance interaction takes
place within two sectors $\varphi_{1}<\varphi<\varphi_{2}$ and
$\varphi_{4}<\varphi<\varphi_{5}.$

For each $\varphi<\varphi_{1}$ the new damping region is a well
separated damping band.  The damping amplitude is small because of
the small factor of order $W^{4}$ mentioned above.  When
$\varphi\to \varphi_{1}-0$, small peak appears near the ``blue''
boundary of this damping region.  This peak is a precursor of the
resonance between the substrate plasmon and the first array QCB
plasmon (Fig.\ref{ImDFAngl-16}).  The same happens from the
opposite side of the sector $(\varphi_{1},\varphi_{2}$ when
$\varphi\to \varphi_{2}+0$ (Fig.\ref{ImDFAngl-33}).

Within the sector $\varphi_{1}<\varphi<\varphi_{2},$ the damping band
contains a well pronounced peak corresponding to resonant interaction
between the substrate plasmon and the first array plasmons (see
Figs.\ref{ImDFAngl-16},\ref{ImDFAngl-33}).  The peak
amplitude is of order of the damping amplitude within the initial
damping region.  It has a Lorentz form placed on a wide and low
pedestal.  The peak is especially sensitive to the strength of the
QCB-substrate interaction which is governed by the distance $D$
between QCB and substrate.

\begin{figure}[h]
\centerline{\epsfig{figure=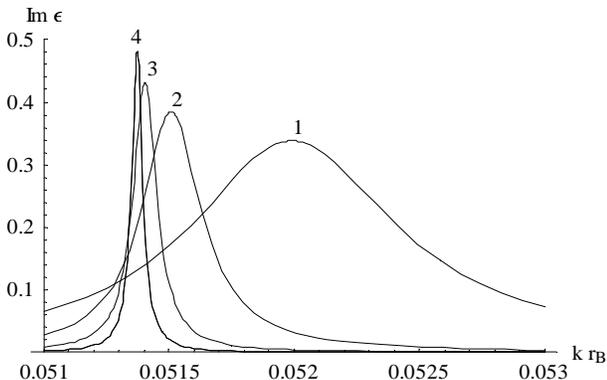,width=80mm,height=50mm,angle=0}}
\caption{Damping tail for $\varphi=20^{\circ}$ for different distances
$D$ between QCB and substrate.  The curves ${\bf{1}}$, ${\bf{2}}$,
${\bf{3}}$, and ${\bf{4}}$ correspond to $D=1$~nm, $1.5$~nm, $2$~nm,
and $2.5$~nm, respectively.  With increasing $D$, the resonant peak
narrows and slowly increases, whereas the area under the curve
decreases.}
\label{ImDFW}
\end{figure}

To study this $D$ dependence, let us consider the imaginary part
$\Im\epsilon({\bf k},\omega)$ of the dielectric function within the
considered sector $\varphi_1<\varphi<\varphi_2.$ In the vicinity of
the plasmon frequency $\omega\approx\omega_s(k)$ this imaginary part
is written as
\begin{eqnarray*}
 \Im\epsilon({\bf k},\omega)=
  \frac{|W_{\bf{k}}|^2}{U_{\bf{k}}}
  \frac{-\Im{w}({\bf{k}},\omega_s(k))}
       {\left(\omega^2-v^2k_1^2\right)^2+
        \left(\Im{w}({\bf{k}},\omega_s(k))\right)^2},
\end{eqnarray*}
with ${w}({\bf{k}},\omega_s(k))$ being of order $|W|^2$.  So the
resonance peak indeed has the Lorentz like shape with height of order
unity,
\begin{eqnarray*}
  \Im\epsilon_{max}\sim
     \frac{|W_{\bf{k}}|^2}
          {|W_{{\bf{k}}-Q{\bf{e}}_2}|^2}\sim1,
\end{eqnarray*}
whereas its half-width
\begin{eqnarray*}
  \Gamma=\Im{w}({\bf{k}},v|k_1|)\sim|W_{{\bf{k}}-Q{\bf{e}}_2}|^2,
\end{eqnarray*}
is of order $W^2$. The peak is displayed in
Fig.\ref{ImDFW} for different values of the distance $D$ between
the substrate and the nearest (first) array. It is seen that the
amplitude changes slowly with increasing distance $D$ while
its width squeezes sharply, $W^2\sim1/D^2$.

There is no resonance interaction within the sector
$\varphi_{2}<\varphi<\varphi_{4}$ but further increase of the angle
$\varphi_{4}<\varphi<\varphi_{5}$ leads to re-appearance of the
resonant peak within the damping tail.  In this case one deals with a
resonance between the substrate plasmon and the QCB plasmon in the
second array.  Existence of this resonance is caused by inter-array
interaction that brings additional small parameter to the imaginary
part of the dielectric function.  As a result, the width of the peak
is much smaller in the second sector while surprisingly, the peak
amplitude has the same order of magnitude as in the case of resonance
with the first array (closest to the substrate).

Existence of the additional QCB bands (tails) of Landau damping
and appearance of the resonant peaks within the bands (tails) is a
clear manifestation of interplay between real $2D$ surface
plasmons and quasi-$2D$ QCB plasmons.

\section{Conclusion}\label{sec:Conclu}

In conclusion, the possibility of spectroscopic studies of the
excitation spectrum of quantum crossbars interacting with
semiconductor substrate is investigated .  A capacitive contact
between QCB and substrate does not destroy the LL character of the
long wave excitations.  However, quite unexpectedly, the interaction
between the surface plasmons and plasmon-like excitations of QCB
essentially influences the dielectric properties of the substrate.
The QCB may be treated as a diffraction grid for the substrate
surface, and Umklapp diffraction processes radically change the
plasmon dielectric losses.  Due to QCB-substrate interaction,
additional Landau damping regions of the substrate plasmons appear.
Their existence, form and density of losses are strongly sensitive to
the QCB period.  So the surface plasmons are more fragile against
interaction with superlattice of quantum wires than the LL plasmons
against interaction with $2D$ electron gas in a substrate.

\section*{Acknowledgments}

The author appreciates numerous helpful discussions with S.
Gredeskul, K. Kikoin, and Y. Avishai.



\end{document}